\documentclass[twocolumn,conference]{IEEEtran} 
\usepackage {epsfig}
\usepackage {latexsym}

\def\BibTeX{{\rm B\kern-.05em{\sc i\kern-.025em b}\kern-.08em
    T\kern-.1667em\lower.7ex\hbox{E}\kern-.125emX}}

\begin{document}

\title{Distributed Algorithms for Computing Alternate Paths Avoiding Failed Nodes and Links}

\author{Amit M. Bhosle\thanks{Currently at Amazon.com, 1200 12$^{th}$ Ave.
S., Seattle, WA - 98144}, Teofilo F. Gonzalez\\
Department of Computer Science\\
University of California\\
Santa Barbara, CA 93106\\
\{bhosle,teo\}@cs.ucsb.edu}

\maketitle

\begin{abstract}
A recent study characterizing failures in
computer networks shows that transient single element 
(node/link) failures are the dominant failures in large
communication networks like the Internet. Thus, having 
the routing paths 
globally recomputed on a failure does 
not pay off since the failed element
recovers fairly quickly, and the recomputed 
routing paths need to be discarded. In this 
paper, we present the first distributed algorithm 
that computes the alternate paths required by 
some {\em proactive recovery scheme}s for handling 
transient failures. Our algorithm computes paths 
that avoid a failed {\em node}, and provides an 
alternate path to a particular destination
from an upstream neighbor of the failed node.
With minor modifications, we can have the 
algorithm compute alternate paths that 
avoid a failed {\em link} as well. To the best 
of our knowledge all previous algorithms proposed 
for computing alternate paths are centralized, and 
need complete information of the network graph 
as input to the algorithm.
\end{abstract}


\section{Introduction}
Computer networks are normally represented by edge weighted graphs.
The vertices represent computers (routers), the edges represent the
communication links between pairs of computers, and the weight of an edge
represents the cost (e.g. time) required to transmit a
message (of some given length) through the link. The links are bi-directional.
Given a computer network represented by
an edge weighted graph $G=(V,E)$, the problem is to find the
best route (under normal operation load)
to transmit a message between every pair of vertices.
The number of vertices ($|V|$) is $n$ and the number of
edges ($|E|$) is $m$.
The shortest paths tree of a node $s$, $\mathcal{T}_s$, specifies the
fastest way of transmitting a message to node $s$ originating
at any given node in the
graph.  Of course, this holds as long as
messages can be transmitted at the specified costs.
When the system carries heavy traffic on some links these routes might not be
the best routes, but under normal operation the routes are the fastest.
It is well known that the all pairs shortest path problem, finding
a shortest path between every pair of nodes, can be computed
in polynomial time.
In this paper we consider the case when the
nodes\footnote{The nodes are
{\em single-} or {\em multi-}processor computers} 
in the network may be susceptible to transient faults.
These are sporadic faults of at most one node
at a time
that last for a relatively short period of time.
This type of situation has been studied in the past 
\cite{bg04jgaa,bg08snfr,lynzc,sl,wg08,znylwc} because it represents
most of the node failures occurring in networks.
{\em Single} node failures represent more than 85\% of all node
failures \cite{mibcd}. Also, these node failures are 
usually {\em transient},
with 46\% lasting less than a minute, and 86\% lasting less than
10 minutes \cite{mibcd}. 
Because nodes fail for relative short periods of time, propagating
information about the failure throughout the network is not recommended.
The reason for this is that it takes time for the information about the
failure to be communicated
to all nodes and it takes time for the nodes to recompute the shortest
paths in order to re-adapt to the new network
environment.
Then, when the failing node recovers, a new messages disseminating
this information needs to be sent to inform the nodes to roll back
to the previous state. This process also consumes resources.
Therefore, propagation of failures is best
suited for the case when nodes fail for long periods of time.
This is not the scenario which characterizes current networks,
and is not considered in this paper.

In this paper we consider the case where the network is {\em biconnected} 
({\em 2-node-connected}), meaning that
the deletion of a single node does not disconnect the network.
Biconnectivity ensures that there is at least one path between every
pair of nodes even in the event that a node fails (provided the failed node
is not the origin or destination of a path).
A ring network is an example of a
biconnected network, but it is not necessary for a network to have
a ring formed by all of its nodes in order to be biconnected.
Testing whether or not a network is biconnected can be performed
in linear time with respect to the number of nodes and links in a network.
The algorithm is based on depth-first search \cite{clrs}.

Based on our previous assumptions about failures, 
a message originating at node $x$ with destination 
$s$ will be sent along the path specified by 
$\mathcal{T}_s$ until it reaches node $s$ or
a node adjacent to a node that has failed.
In the latter case, we need to use a
recovery path to $s$ from that point.
Since we assume single node faults and
the graph is biconnected, such a path always exists. 
We call this problem of finding the recovery paths the
{\em Single Node Failure Recovery (SNFR)} problem.
In this paper, we present an efficient distributed
algorithm to compute such paths. Also, our algorithm 
can be generalized to solve some other problems
related to finding alternate paths or edges.

A distributed algorithm for computing the alternate
paths is particularly useful if the routing tables
themselves are computed by a distributed 
algorithm since it takes away the need 
to have a centralized view of the entire
network graph. Centralized algorithms 
inherently suffer from the
overhead on the network administrator to put together
(or source and verify) a consistent snapshot 
of the system, in order to feed it to the algorithm. 
This is followed by the need to deploy the output
generated by the algorithm (e.g. alternate path 
routing tables) on the relevant computers (routers) 
in the system.  Furthermore, centralized algorithms 
are typically resource intensive since a single computer 
needs to have enough memory and processing power 
to process a potentially 
huge network graph. Some other advantages 
of a distributed algorithm are reliability 
(no single points of failure), scalability and 
improved speed (computation time).

\subsection{Related Work}

A popular approach of tackling the issues related to transient
failures of network elements is that of using {\em proactive recovery
schemes}. These schemes typically work by precomputing alternate
paths at the network setup time for the failure scenarios, 
and then using these alternate
paths to re-route the traffic when the failure actually occurs.
Also, the information of the failure is suppressed
in the hope that the failure is transient and the failed element
will recover shortly. The local 
rerouting based solutions proposed in 
\cite{bg08snfr,lynzc,sl,wg08,znylwc} fall into this category.

Zhang, et. al. \cite{znylwc} present protocols based on 
local re-routing for dealing with transient single node failures. 
They demonstrate via simulations that the recovery paths computed 
by their algorithm are usually within 15\%
of the theoretically optimal alternate paths.

Wang and Gao's Backup Route Aware Protocol 
(BRAP) \cite{wg08} also uses some precomputed backup 
routes in order to handle transient
single {\em link} failures. One problem central to their 
solution asks for the availability of {\em reverse paths} 
at each node. However, they do not discuss the computation 
of these reverse paths. As we discuss later, the alternate 
paths that our algorithm computes qualify as
the reverse paths required by the BRAP protocol 
of \cite{wg08}.

Slosiar and Latin \cite{sl} studied the single {\em link} failure
recovery problem and presented 
an $O(n^3)$ time for computing the link-avoiding alternate
paths. A faster algorithm, with a running time of $O(m + n\log n)$
for this problem was presented in \cite{bg04jgaa}. The local-rerouting 
based fast recovery protocol of \cite{bg08snfr} 
can use these paths to recover from 
single link failures as well. Both these algorithms, \cite{bg04jgaa,sl},
are centralized algorithms that work using the information
of the entire communication graph. 

\subsection{Preliminaries}
\label{prelim}

Our communication network is modeled by an edge-weighted 
biconnected undirected 
graph $G=(V,E)$, with $n=|V|$ and $m=|E|$.
Each edge $e\in E$ has an 
associated cost (weight), denoted by $cost(e)$, which 
is a non-negative real number.
We use $p_G(s,t)$ to denote a shortest path between $s$ and $t$
in graph $G$ and $d_G(s,t)$ to denote its cost. 

A shortest path tree $\mathcal{T}_s$ for a node 
$s$ is a collection of
$n-1$ edges $\{e_1,e_2,\ldots,e_{n-1}\}$ of $G$
which form a spanning tree of $G$ such
that the path from node $v$ to $s$ in $\mathcal{T}_s$
is a shortest path from $v$ to $s$ in $G$.
We say that $\mathcal{T}_s$ is rooted at node $s$. With
respect to this root we define the set of nodes that are
the {\em children} of a node $x$ as follows.
In $\mathcal{T}_s$ we say that every node $y$ 
that is adjacent to $x$ such
that $x$ is on the path in $\mathcal{T}_s$ from 
$y$ to $s$, 
is a child of $x$. For each node $x$ in the shortest 
paths tree, $k_x$ denotes the number of 
children of $x$ in the tree, and 
$\mathcal{C}_x = \{x_1, x_2, \ldots x_{k_x}\}$ 
denotes this set of children of the node $x$. 
Also, $x$ is said to be the {\em parent} of each 
$x_i\in \mathcal{C}_x$ in the tree $\mathcal{T}_s$.
The parent node, $p$, of a node $c$ is sometimes 
referred to as a {\em primary neighbor} or 
{\em primary router} of $c$, while $c$ is
referred to as an {\em upstream neighbor} or
{\em upstream router} of $p$. The children of
a particular node are said to be {\em siblings}
of each other.

$V_x(\mathcal{T})$ denotes the set of nodes 
in the subtree of $x$ in the tree $\mathcal{T}$ 
and $E_x\subseteq E$ denotes the set of all 
edges incident on the node $x$ in the graph $G$. 
$nextHop(x,y)$ denotes the next node from 
$x$ on the shortest path from $x$ to $y$.
Note that by definition, $nextHop(x,y)$ is the 
parent of $x$ in $\mathcal{T}_y$.

\subsection{Problem Definition}
\label{def}
The Single Node Failure Recovery problem is formally 
defined in \cite{bg08snfr} as follows:

{\tt SNFR}: Given a biconnected undirected edge 
weighted graph $G=(V,E)$,
and the shortest paths tree 
$\mathcal{T}_s(G)$ of a node $s$ in $G$ where 
$\mathcal{C}_x=\{x_1, x_2, \ldots x_{k_x}\}$
denotes the set of {\em children} of $x$ in 
$\mathcal{T}_s$, for each node $x \in V$ and $x \neq s$, find 
a path from $x_i \in \mathcal{C}_x$ to $s$ in the graph 
$G=(V\setminus \{x\}, E\setminus E_x)$, where $E_x$ is the
set of edges adjacent to $x$.

In other words, for each node $x$ in the graph, we 
are interested in finding alternate paths from each 
of its children in $\mathcal{T}_s$ to the node $s$ 
when the node $x$ {\em fails}. Note that the problem 
is not well defined when node $s$ fails.

The above definition of alternate paths matches 
that in \cite{wg08} for {\em reverse paths}: for each 
node $x\in G(V)$, find a path from $x$ to the node 
$s$ that does not use the primary neighbor (parent node) 
$y$ of $x$ in $\mathcal{T}_s$.

\subsection{Main Results}
\label{main-results}

Our main result is an efficient distributed 
algorithm for the SNFR problem. Our 
algorithm requires $O(m + n)$ messages 
to be transmitted among the nodes (routers), 
and has a space complexity of $O(m+n)$ across 
{\em all} nodes in the network (this, being 
asymptotically equal to the size of the entire 
network graph, is asymptotically {\em optimal}). 
The space requirement at any single node is
linearly proportional to the number of children 
(the node's degree) and the number 
of siblings that the node has in 
the shortest paths tree of the destination $s$. 
When used for multiple {\em sink} nodes in the 
network, the space complexity at each node is
bounded by its total number of children and
siblings across the shortest paths trees of
all the sink nodes. Note that even though 
this is only bounded by $O(n^2)$ in theory (since 
each node in the network can be a sink, and
a node can theoretically have $O(n)$ children), 
it is much smaller in practice ($O(n)$: for $n$
sink nodes, as average node degree in shortest 
paths trees is usually within 20-40 even for 
$n$ as high as a few $1000$s). Finally, we 
discuss the scalability issues that may occur 
in large networks. 

Our algorithm is based on a request-response
model, and does not require any {\em global coordination}
among the nodes.

To the best of our knowledge, this is the first completely
decentralized and distributed algorithm for computing alternate
paths. All previous algorithms, including those presented in 
\cite{bg04jgaa,bg08snfr,lynzc,sl,wg08,znylwc} are centralized algorithms 
that work using the information of the entire network graph 
as input to the algorithms.

Furthermore, our algorithm can be generalized to solve other similar
problems. In particular, we can derive distributed algorithms for: the
single link failure recovery problem studied in \cite{bg04jgaa,sl}, minimum
spanning trees sensitivity problem \cite{dixon} and 
the detour-critical edge problem \cite{npw-detour}. The cited papers 
present centralized algorithms for the respective problems. 

\section{Key Properties of the Alternate Paths}
\label{key-properties}
We now describe the key properties of the alternate paths 
to a particular destination that can be used by a node 
in the event of its parent node's failure. These same
principles have been used in the design of the centralized
algorithm in \cite{bg08snfr}. However, for completeness, we
discuss them briefly here.

\begin{figure}[h]
\centerline{\epsfig{file=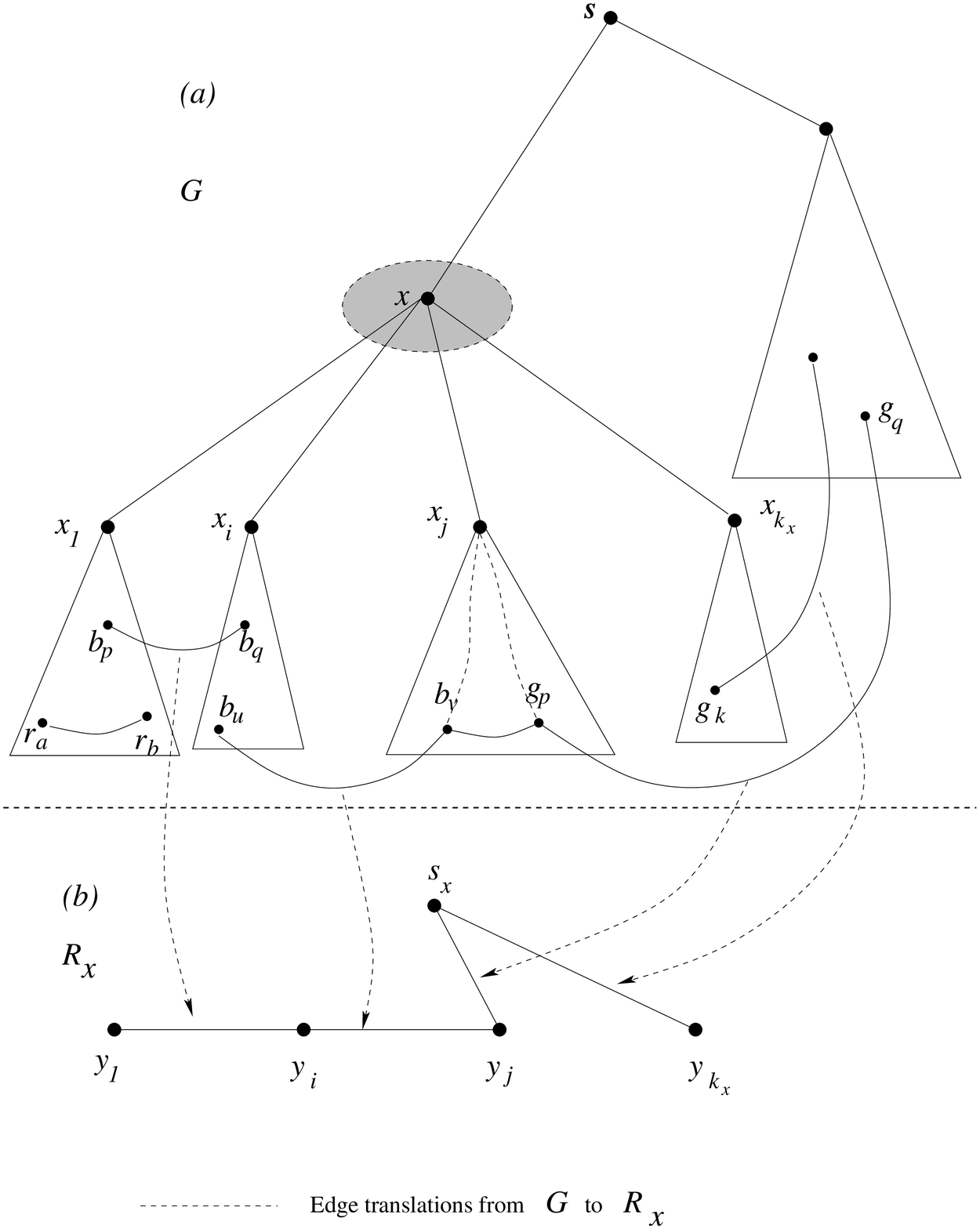 , width = 1.0\linewidth}}
\caption{
Recovering from the failure of $x$: 
Constructing the recovery graph $\mathcal{R}_x$}
\label{main}
\end{figure}

Figure \ref{main}(a) illustrates a scenario of a single node failure. 
In this case, the node $x$ has failed, and we need to find alternate 
paths to $s$ from each $x_i\in \mathcal{C}_x$. When a node fails, 
the shortest paths tree of $s$, $\mathcal{T}_s$, gets split into 
$k_x + 1$ components - one containing the source node $s$ and each 
of the remaining ones containing the subtree of a child 
$x_i \in \mathcal{C}_x$.

Notice that the edge $\{g_p, g_q\}$,
which has one end point in the subtree of $x_j$, and the other 
outside the subtree of $x$ provides a candidate recovery path 
for the node $x_j$. The complete path is of the form 
$p_G(x_j, g_p) \leadsto \{g_p, g_q\} \leadsto p_G(g_q, s)$. 
Since $g_q$ is outside the subtree of $x$, the path $p_G(g_q, s)$ 
is not affected by the failure of $x$. Edges of this type 
(from a node in the subtree of $x_j \in \mathcal{C}_x$ to a 
node outside the subtree of $x$) can be used by 
$x_j \in \mathcal{C}_x$ to {\em escape} the failure of node 
$x$. Such edges are called {\em green} edges.  For example,
the edge $\{g_p, g_q\}$ is a green edge.

Next, consider the edge $\{b_u, b_v\}$
between a node in the subtree of $x_i$ and a node in the subtree 
of $x_j$. Although there is no green edge with an end point in 
the subtree of $x_i$, the edges $\{b_u, b_v\}$ and $\{g_p, g_q\}$
together offer a candidate recovery path that can be used by 
$x_i$ to recover from the failure of $x$. Part of this path 
connects $x_i$ to $x_j$ 
($p_G(x_i, b_u) \leadsto \{b_u, b_v\} \leadsto p_G(b_v, x_j)$), 
after which it uses the recovery path of $x_j$ (via $x_j$'s 
green edge, $\{g_p, g_q\}$). Edges of this type (from a node 
in the subtree of $x_i$ to a node in the subtree 
of a sibling $x_j$ for some $i \not= j$) are called 
{\em blue} edges. $\{b_p, b_q\}$ is another blue edge 
and can be used by the node $x_1$ to recover from the 
failure of $x$.

Note that edges like $\{r_a, r_b\}$ and $\{b_v, g_p\}$ 
with both end points within the subtree of the
same child of $x$ do not help any of the nodes in $\mathcal{C}_x$
to find a recovery path from the failure of node $x$.
We do not consider such {\em red} edges 
in the computation of recovery paths,
even though they may provide a shorter recovery path for some nodes
(e.g. $\{b_v,g_p\}$ may offer a shorter recovery path to $x_i$).
The reason for this is that routing protocols 
would need to be quite complex in order to use this 
information. As we describe later in 
the paper, we carefully organize the {\em green} and {\em blue} 
edges in a way that allows us to retain only these edges and 
eliminate useless (red) ones efficiently.

We now describe the construction of a new graph $\mathcal{R}_x$,
called the {\em recovery graph} of $x$, which
will be used to compute recovery paths for the elements of 
$\mathcal{C}_x$ when the node $x$ fails. A
single source shortest paths computation on this graph 
suffices to compute the recovery paths for all 
$x_i \in \mathcal{C}_x$.

The graph $\mathcal{R}_x$ has $k_x + 1$ 
nodes, where $k_x = |\mathcal{C}_x|$. A special node, 
$s_x$, represents in $\mathcal{R}_x$, the node $s$ in the 
original graph $G=(V,E)$. Apart from $s_x$, we have
one node, denoted by $y_i$, for each $x_i\in \mathcal{C}_x$.
We add all the {\em green} and {\em blue} edges defined earlier 
to the graph $\mathcal{R}_x$ as follows. 
A green edge with an end point 
in the subtree of $x_i$ (by definition, green edges have the 
other end point outside the subtree of $x$) translates 
to an edge between $y_i$ and $s_x$. A blue edge with an end 
point in the subtree of $x_i$ and the other in the subtree of 
$x_j$ translates to an edge between nodes $y_i$ and $y_j$.

Note that the weight of the edges added to $\mathcal{R}_x$ 
need not be the same as the weight of the 
corresponding green or blue edges in $G=(V,E)$. 
The weights assigned to the edges in 
$\mathcal{R}_x$ should take into account the weight 
of the actual subpath in $G$ corresponding to the edge
in $\mathcal{R}_x$. 
As long as the weights of edges in $\mathcal{R}_x$ 
don't change with $x$, or can be determined
locally by the node, they can be directly used in our
algorithm. The candidate recovery path of $x_j$ that uses the 
green edge $e = \{u,v\}$ has total cost given by:

\begin{equation}
\label{green-weight}
greenWeight(e) = d_G(x_j, u) + cost(u, v) + d_G(v, s)
\end{equation}

This weight captures the weight of the actual subpath in $G$
corresponding to the edge added to $\mathcal{R}_x$.
However, since the weight given by equation 
(\ref{green-weight}) for an edge depends on the node $x_j$ whose 
recovery path is being computed, it will typically be different
in each $\mathcal{R}_x$ in which $e$ appears as a green edge.
The following weight function is more efficient since it
remains constant across all $\mathcal{R}_x$ graphs that 
$e$ is part of. 

\begin{eqnarray}
\label{fixed-green-weight}
\lefteqn{greenWeight(e)}\nonumber \\
&&= d_G(s,x_j) + d_G(x_j,u) + cost(u,v) + d_G(v,s) \nonumber\\
&&= d_G(s,u) + cost(u,v) + d_G(v,s)
\end{eqnarray}

Note that the correct weight (as defined by 
equation (\ref{green-weight})) to be used for an $\mathcal{R}_x$
can be derived by the node $x$ from the weight 
function defined above by subtracting 
$d_G(s,x_j) = d_G(s,x) + cost(x,x_j)$. Also, the green edge
with an end point in the subtree of $x_j$ with the minimum 
$greenWeight$ remains the same, immaterial of the greenWeight
function (equations (\ref{green-weight}) or 
(\ref{fixed-green-weight})) used since equation 
(\ref{fixed-green-weight}) basically adds the value 
$d_G(s,x_j)$ to all such edges.

As discussed earlier, a blue edge provides a path connecting
two siblings of $x$, say $x_i$ and $x_j$. Once the path reaches 
$x_j$, the remaining part of the recovery path of $x_i$ 
coincides with that of $x_j$. If $b=\{p, q\}$ is the blue 
edge connecting the subtrees of $x_i$ and $x_j$ 
the length of the subpath from $x_i$ to $x_j$ is:

\begin{equation}
\label{blue-weight}
blueWeight(b) = d_G(x_i, p) + cost(p, q) + d_G(q, x_j)
\end{equation}

We assign this weight to the edge corresponding to the blue edge
$\{p, q\}$ that is added in $\mathcal{R}_x$ between $y_i$ 
and $y_j$. 

Note that if $w$ is the nearest common ancestor
of the two end points $u$ and $v$ of and edge $e=(u,v)$, $e$
is a green edge in the $\mathcal{R}$ graphs for all nodes
on path between $w$ and $u$, and $w$ and $v$ (excluding $u$,
$v$ and $w$: it is a blue edge in $\mathcal{R}_w$, and is
unusable in $\mathcal{R}_u$ and $\mathcal{R}_v$ since a node
$z$ is deemed to have failed while constructing
$\mathcal{R}_z$). Assuming that a node can determine 
whether an edge is blue or green in its recovery graph (we
discuss this in detail in the next section), it
is easy to see that it can derive the edge's blue weight
from its green weight:
\begin{eqnarray}
\lefteqn{blueWeight(e) = greenWeight(e) - }\nonumber \\
&&(2\cdot d_G(s,w) + cost(w,w_u) + cost(w,w_v))
\end{eqnarray}
where $w_u$ and $w_v$ are respectively the child nodes of $w$ 
whose subtrees contain the nodes $u$ and $v$. Information about
all terms being subtracted is available locally at $w$, and
consequently, the greenWeight and blueWeight values for
an edge can be computed/derived using information local 
to the node $w$.

If there are multiple green edges 
with an end point in $V_{x_j}$, the subtree of $x_j$,
we choose the one which offers the shortest recovery path for
$y_j$ (with ties being broken arbitrarily) and ignore the rest.
Similarly, if there are multiple edges between the subtrees of
two siblings $x_i$ and $x_j$, we retain the one which offers
the cheapest alternate path.

The construction of our graph $\mathcal{R}_x$ is now complete.
Computing the shortest paths tree of $s_x$ in $\mathcal{R}_x$ provides 
enough information to compute the recovery paths for all nodes 
$x_i \in \mathcal{C}_x$ when $x$ fails. 

Note that any edge $e=(u,v)$ acts as a blue edge in at most one
$\mathcal{R}_x$: that of the nearest-common-ancestor of $u$ and $v$. 
Also, any node $c\in G(V)$ belongs to exactly one $\mathcal{R}_x$:
that of its parent in $\mathcal{T}_s$. As we discuss later,
the space requirement at any node is 
linearly proportional to the number of children and the 
number of siblings that it has.

Figure \ref{main} illustrates the consturction of $\mathcal{R}_x$ used to
compute the recovery paths from the node $x_i\in \mathcal{C}_x$ 
to the node $s$ when the node $x$ has failed. In this simple
example, the path from $y_i$ to $s_x$ is 
$y_i \leadsto y_j \leadsto s_x$. The corresponding recovery path for
$x_i$ is $p_G(x_i, b_u) \leadsto \{b_u, b_v\} \leadsto p_G(b_v,x_j)$, 
followed by the recovery path of $x_j$: 
$p_G(x_j, g_p) \leadsto \{g_p, g_q\} \leadsto p_G(g_q, s)$.

\section{A Distributed Algorithm for Computing the Alternate Paths}
\label{distributed}

In this section, we use the basic principals of
the alternate paths described earlier to design an efficient
distributed algorithm for computing the alternate paths.

\subsection{Computing the DFS Labels}

Our distributed algorithm requires that each node in the
shortest paths tree $\mathcal{T}_s$ maintain its $dfsStart(\cdot)$
and $dfsEnd(\cdot)$ labels in accordance with how a
depth-first-search (DFS) traversal of $\mathcal{T}_s$ starts
or ends at the node. Ref. \cite{fppp} reports 
efficient distributed algorithms for this particular problem 
(of assigning lables to the nodes in a tree as dictacted
by a DFS traversal of the tree). The
basic algorithm reported in Ref. \cite{fppp},
named {\tt Wake \& Label$_A$}, assigns DFS
labels to the nodes in the range $[1,n]$ 
in asymptotically optimal time and requires 
$3n$ messages to be exchanged between the
nodes. They also discuss other variations of this
algorithm which vary with respect to the time required
to assign the labels, the range of labels, and the
number of messages exchanged between the nodes in
the network. An appropriate algorithm can be chosen
to assign the $dfsStart(\cdot)$ and $dfsEnd(\cdot)$
labels required for our distributed algorithm.

We sketch below the basic algorithm, {\tt Wake \& Label$_A$}
below.

The {\tt Wake \& Label$_A$} algorithm runs in three
phases: {\em wakeup}, {\em count}, and {\em allocation}.
In the first ({\em wakeup}) phase, which is a top-down phase,
the root node sends
a message to all of its child nodes asking them to report
the number of nodes in their subtree (including themselves).
The child nodes recursively pass on the message to their
children. In the second ({\em count}) phase,
which is a bottom-up phase, each node reports the size of
its subtree to its parent node. The variants of the
{\tt Wake \& Label} algorithms differ in the last phase
({\em allocation}) which deals with assigning the labels
to the nodes of the tree. In the simplest version, once
the root node knows the value of $n$ (the total number
of nodes in the tree), knowing the size of the subtrees
of each child node, it can split the range $[1,n]$
disjointly among its children, and each child node 
recursively assigns a sub-range to its children (a
child with $c$ nodes in its subtree is assigned a
range containing $c$ values).

The reader is referred to Ref. \cite{fppp} for the
detailed description and analysis of the {\tt Wake 
\& Label$_A$} algorithm and its variants. For computing 
the $dfsStart(\cdot)$ and $dfsEnd(\cdot)$ labels
required by our algorithm, the total range of these
labels across all the nodes in $\mathcal{T}_s$ is
$[1,2n]$, and a child with $c$ children is assigned
a range of $2c$ values. All other aspects of any
of the DFS label assignment algorithms reported in 
Ref. \cite{fppp} can be used as appropriate.
Note that even though it is not explicitly 
mentioned in Ref. \cite{fppp}, the 
{\tt Wake \& Label$_A$} algorithm (including 
our modifications) can be implemented on a 
request-response model, without the need 
of any global clock for coordination across 
the nodes.

\subsection{Collecting the Green and Blue Edges}

Our algorithm requires that each node in the 
network maintain the following data-structures:

1. {\tt ParentBlueEdges List}: The list of edges in 
the network graph which have 
one end point within the subtree of the node, and the 
other end point in the subtree of a sibling node. I.e. 
all edges from the node's subtree that are {\em blue} 
in the recovery graph $\mathcal{R}$ of the node's parent.

2. {\tt ChildrenGreenEdges Map}: A map that stores for 
each child node, the cheapest green edge with 
an end point in the child node's subtree. Recollect 
that a green edge of a node has the other end 
point outside the subtree of the node's parent.

We now discuss the details of this part of the 
algorithm for building the {\tt ParentBlueEdges} 
and {\tt ChildrenGreenEdges} data-structures. 
A procedure, {\tt CollectNonTreeEdges}, 
triggers a protocol where 
each node recursively asks each of its 
children to forward it the non-tree edges 
that have an end point in the child's subtree. 
Each node processes all its own non-tree 
edges, and those forwarded by a child node.
For processing a non-tree edge, a node uses
the $dfsStart(\cdot)$ and $dfsEnd(\cdot)$ labels
of the edge's two end points to decide whether
the edge should be added to its {\tt ParentBlueEdges}
list or the {\tt ChildrenGreenEdges} map. 
For an edge to be added to the {\tt ParentBlueEdges} 
list, the edge should
have exactly one end point in the node's subtree, 
while the other end point still be within the
parent's subtree (but outside this node's subtree).
For each edge that is forwarded by a child,
the node updates the corresponding entry for the
child in the {\tt ChildrenGreenEdges} map if
the newly forwarded edge is cheaper than the
edge currently stored for the child.
Finally, if at least one of the two end points 
of the edge lies outside this
node's subtree, it forwards the information of the 
edge to the parent after updating its local 
data-structures. Otherwise, it simply discards 
the edge and does not forward it to its parent.
The reason for this is that edges whose both end
points belong to a node's subtree cannot serve as
a blue or green edge in the recovery graph of
the node's parent, and informing the parent about
such an edge does not serve any purpose (if this 
node is the nearest-common-ancestor of the
edge's two end points, the edge would be stored in
the {\tt ParentBlueEdges} lists at the two child
nodes whose subtrees contain the edge's end points).

A child node invokes the proceudre {\tt RecordNonTreeEdge} 
defined below on its parent, with a message $\mathcal{M}$ 
containing the following information associated with a 
non-tree edge $e$:
\begin{itemize}
\item $e=(p_1, p_2)$: The non-tree edge, with 
   $p_1$ and $p_2$ as the end points.
\item $weight(e)$: Weight of the edge $e$.
\item $senderId$: Id of this child node 
   sending the message to the parent node.
\end{itemize}
These individual pieces,
$e$, $p_1$, $p_2$, and $senderId$, can respectively be accessed via 
$\mathcal{M}$ using the methods 
$\mathcal{M}${\tt.edge},
$\mathcal{M}${\tt.$p_1$},
$\mathcal{M}${\tt.$p_2$} and 
$\mathcal{M}${\tt.senderId}.

\vspace*{0.1in}
\noindent
\hspace*{0.0in} {\tt \bf Procedure RecordNonTreeEdge($\mathcal{M}$)}\\
\hspace*{0.1in} {\tt if (isMyDescendant($\mathcal{M}$.$p_1$) AND}\\
\hspace*{0.18in} {\tt isMyDescendant($\mathcal{M}$.$p_2$)) do:} \\
\hspace*{0.18in} {\tt // both end points in my }\\
\hspace*{0.18in} {\tt // subtree: ignore}\\
\hspace*{0.18in} {\tt return;}\\
\hspace*{0.1in} {\tt fi}\\
\hspace*{0.1in} {\tt // retrieve the current green }\\
\hspace*{0.1in} {\tt // edge for this sender from }\\
\hspace*{0.1in} {\tt // the ChildrenGreenEdges map }\\
\hspace*{0.1in} {\tt Edge existing = }\\
\hspace*{0.18in} {\tt CGE.get($\mathcal{M}$.senderId);}\\
\hspace*{0.1in} {\tt Edge edge = $\mathcal{M}$.edge;}\\
\hspace*{0.1in} {\tt if (existing == null OR }\\
\hspace*{0.18in} {\tt edge.weight < existing.weight), do: }\\
\hspace*{0.18in} {\tt // if new or cheaper edge, }\\
\hspace*{0.18in} {\tt // update our data-structure }\\
\hspace*{0.18in} {\tt CGE.put($\mathcal{M}$.senderId, edge);}\\
\hspace*{0.1in} {\tt fi}\\
\hspace*{0.1in} {\tt if (edgeIsBlueForParent(edge)), do:}\\
\hspace*{0.15in} {\tt ParentBlueEdges.add(edge);}\\
\hspace*{0.1in} {\tt fi}\\
\hspace*{0.1in} {\tt // Reset the senderId, }\\
\hspace*{0.1in} {\tt // and forward edge to parent}\\
\hspace*{0.1in} {\tt $\mathcal{M}$.senderId = self.id; }\\
\hspace*{0.1in} {\tt parent.RecordNonTreeEdge($\mathcal{M}$); }\\
\hspace*{0.0in} {\tt \bf End RecordNonTreeEdge}\\

The {\tt edgeIsBlueForParent} method used above determines
whether or not an edge is blue for this node's parent. This
can be determined easily if the node knows its parent's 
$dfsStart(\cdot)$ and $dfsEnd(\cdot)$ labels. For efficiency,
after the DFS labels have been computated,
each node can query its parent for its labels, and
store these locally. In some cases, these values can just be
queried from the parent node as and when needed.

\subsection{Computing the Alternate Paths to Recover 
from a Node's Failure}

Once the edge propagation phase is over, part of the information
required to construct $\mathcal{R}_x$, the recovery graph of $x$,
is available at the node $x$, and the remaining is available at
the children of $x$. In particular, $x$ has the information about
the nodes of $\mathcal{R}_x$ and the green edges of 
$\mathcal{R}_x$, while the children of $x$ have the information
of the blue edges of $\mathcal{R}_x$. 

Conceptually, $x$ can construct the entire graph $\mathcal{R}_x$
locally, and compute the shortest paths tree of $s_x$. This process
would result in a space complexity of 
$O(m_x + n_x)$ at node $x$, where $m_x$ 
and $n_x$ denote the number of edges and 
nodes in $\mathcal{R}_x$ respectively.
Note that $m_x$ can be as large as $O(n_x^2)=O(|\mathcal{C}_x|^2)$.
In order to keep the space requirement low, the shortest paths tree,
$\mathcal{T}_{s_x}$, of $s_x$ is built incrementally, by looking
at the edges of $\mathcal{R}_x$ only when they 
are needed. Essentially, we use
the edges exactly in the order dictated by the Dijkstra's shortest
paths algorithm\cite{dijk}. $x$ initially builds $\mathcal{R}_x$
using the information it locally has: the $k_x+1$ nodes, and the
green edge from $y_i$ to $s_x$ for $1\leq i \leq k_x$ (if the
{\tt ChildrenGreenEdges} map has an entry for $x_i$). $x$ maintains
a priority queue data structure, {\tt candidates}, 
which initially has an
entry for each $y_i$, with a priority\footnote{lower value
implies higher priority} equal to the weight of the edge 
between $s_x$ and $y_i$\footnote{if no edge is present, a 
priority of $\infty$ is assigned}. The remaining steps of the
algorithm are as follows.
\begin{enumerate}
\item While there are more entries in {\tt candidates}, execute
steps 2 - 4.
\item Delete entry from {\tt candidates} with highest priority.
\item Assign the priority value as the final distance (from $s_x$)
for the node $y_p$ associated with the queue entry.
\item Fetch the blue edges from child node $x_p$. For each blue edge
thus retrieved, if it provides a shorter path to its other end point,
say $x_q$, update the priority of the queue entry 
corresponding to $y_q$ with this value.
\end{enumerate}

Note that the blue edges stored at a child node $x_p$ are retrieved
only when they are needed by the algorithm, and that each node $x$
needs space linearly proportional to its number of children, and the
number of its siblings. For each sibling, a node needs to store at most
one edge (which has the smallest blue weight) with an end point 
in its own subtree, and the other in the sibling's subtree. These
edges are the blue edges that are added to the parent node's recovery
graph. Using Fibonacci heaps\cite{fib} for the priority queue, 
$\mathcal{T}_{s_x}$ can be computed in $O(m_x + n_x\log n_x)$ time.

\section{Scalability Issues}
\label{scalability}

In large communication networks, the nodes at higher levels in the 
shortest paths tree (i.e. {\em closer} to the destination) 
may face scalability issues. This happens 
primarily because such nodes have large subtrees, and consequently
a large number of edges may have an end point in their subtrees.
Receiving information about all these edges may potentially
overwhelm the nodes. In this section, we discuss a few approaches
to deal with such issues. The applicability of the 
approaches varies with the particular network topology, and the
resources (mainly, the amount of temporary storage) available 
at the routers.

\subsection*{Producer Consumer Problem}
The problem of a node receiving the information of edges from its
child nodes, and processing this information can be considered to
be a {\em producer-consumer} problem, where the child nodes
{\em produce} the edges, and a parent node {\em consumes} the
edge by processing it. The scalability issues occur in a case 
where all the child nodes together attempt to 
deliver the edges to their parent at a rate higher
than the rate at which the parent node can process the edges.
Recollect that processing an edge by a node includes updating 
its local data structures (if applicable), and delivering the
information of the edge to the parent node.

Our approaches of dealing with these scalability issues can be
categorized in two broad categories: (a) The consumer tries to
minimize the processing time (and thus, increase the consumption
rate), and (b) the producers co-ordinate among themselves to 
limit the rate at which the consumer receives the information
to be consumed.

\subsection*{Consumer Driven Solutions}

The key principals of this approach are the following. 
(a) If a parent node is too busy to process 
a new edge, it can {\em reject} the delivery attempt of the edge
by the child node. For the parent node, a 
rejected delivery is equivalent to no delivery attempt 
at all. (b) For a child node whose attempt 
to deliver an edge was rejected by its parent, 
the {\em processing} of the edge is still incomplete. 
To complete the processing, it {\em must} successfully
deliver the edge to the parent. For a rejected delivery, the
node must {\em retry} the deliver some time in future.

The fact that a node may need to retry the delivery of an edge to
its parent essentially translates to the requirement that the
node have access to a temporary storage space where it can
store the edges whose deliveries were rejected by its 
parent. Otherwise, the delivery of the edge will need
to be {\em transitively} rejected by all nodes down to the node
that initiated the edge's delivery the very first time. Such
options are usually prohibitively expensive, since blips in the
network could also result in an edge not being successfully 
delivered to a parent node. After the edge has been 
successfully delivered to the parent, its corresponding 
entry can be deleted from the temporary storage.

The temporary storage space can be either local 
or remote storage, depending on the size of the
network, and the hardware configuration of the routers. Using
the temporary storage, we split the {\em receipt}, and 
{\em processing}  of an edge into two independent parts. 
As part of receiving an edge, the parent node just needs 
to store the edge into the temporary storage. Once it has 
successfully stored the edge, it acknowledges the delivery
attempt of the child node. Next, each node runs a processing 
{\em daemon}, which reads the information persisted in the
temporary storage and processes the edges. The last step of
this processing includes successfully delivering the information
of the edge to the node's parent. After successful delivery,
the information about the edge from the temporary storage is
deleted. In case the delivery is rejected, the edge is kept
in the storage, and its delivery is retried after some time.

Remote storage solutions could also be used as the temporary
storage space. In particular, the Simple Queue Service (SQS), 
offered by Amazon Web Services \cite{aws} is very well
suited for this use case. The SQS is a highly available
and scalable web service, which exposes a {\em queue}
interface via web service APIs. The APIs of our interest are
{\tt enqueue(Message)}, {\tt readMessage()} and 
{\tt dequeue(MessageId)}. Note that although SQS is not 
a free service, its {\em pay-as-you-go} 
usage-based pricing model makes it a cheaper 
alternative to the traditional option of having large 
hard disks on the routers (and especially more attractive
for this use case since the temporary storage space is
required only during the network set-up time). 
Also, it essentially provides
an {\em unlimited} storage space since there's no 
restriction on the number of messages that can be 
stored in an SQS instance, and can thus be used immaterial
of the network size. When used in our protocol, each node
instantiates an SQS instance for itself, and uses it as its
temporary storage space.

\subsection*{Producer Driven Solutions}

The second approach that we discuss here is based on the
producers co-ordinating amongst themselves to limit the 
rate at which the consumer receives the information
to be consumed.

For simplicity, we assume that the number of edges 
with an end point in the subtree of a node $x_i$ 
(and which need to be forwarded to its parent $x$) is 
proportional to the size of the subtree $V_{x_i}$.
If all the nodes $x_i$ for 
$1 \leq i \leq |\mathcal{C}_x|$
can coordinate amongst themselves about their edge 
deliveries to $x$, they can, to a certain extent, 
ensure that node $x$ does not receive information
about all the edges in a very short window of time.
Essentially, a node $x_k$ is assigned a total time 
proportional to $|V_{x_i}| / |V_x|$ for delivering
its edges to the parent $x$, in order to ensure
that a child node is assigned enough time to deliver
all of its edges to $x$.

Note that this approach relies on the ease
of achieving coordination among all the child nodes
of a node about delivering the edges.

\section{Other Routing Path Metrics}
Though the shortest paths metric is a popular metric
used in the selection of paths, several networks use
some other metrics to select a preferred path. Examples
include metrics based on link bandwidth, network delay, 
hop count, load, reliability, and communication cost. 
Ref. \cite{bhsw} presents a survey on the popular
routing path metrics used. It is interesting to note
that some of these metrics (e.g. communication cost,
hop-count) can be translated to shortest path metrics.
Optimizing hop-count is same as computing shortest
paths where all edges have the same (1 unit) weight,
while communication cost can be directly used as 
edge weights. For optimizing metrics 
like path {\em reliability} 
and {\em bandwidth}, the shortest path algorithms
can be used with easy modification (e.g. the 
reliability of an entire 
path is the product of the reliabilities of the
individual edges; the bandwidth of a path is the
minimum bandwidth across the individual edges on 
the path). For these metrics, algorithms based on 
shortest paths can be directly used with the 
appropriate modifications.

A minimum spanning tree, which constructs a 
spanning tree with minimum total weight is also
used in some networks when the primary goal is to
achieve {\em reachability}.

Note that although we discuss our algorithm in context
of shortest paths, the techniques can be generalized
to find alternate paths in accordance with other 
metrics, and our algorithm can be used with appropriate
modifications. 

The modifications required would be in the weight
functions (Equations \ref{green-weight},
\ref{blue-weight}) used for assigning weights to the edges
added to $\mathcal{R}_x$, the recovery graph that
is constructed to find alternate paths when the node
$x$ fails. Furthermore, paths in $\mathcal{R}_x$
should be computed as dictated by the metric. E.g.
constructing a minimum spanning tree of $\mathcal{R}_x$,
or finding a maximum bandwidth path, etc. 
It is important to note that the process of constructing
$\mathcal{R}_x$ can be modified so that it contains
information about a wide variety of alternate paths that
avoid the failed node $x$ and are relevant for the
particular metric being optimized. An appropriate 
alternate path can be constructed depending on the 
metric of interest, and other factors that affect
path selection.

In large networks, nodes typically denote 
autonomous systems (AS), which are networks owned
and operated by a single administrative entity.
It is common for the paths to be selected based on
inter-AS policies. See Ref. \cite{cr-policy} for a
detailed discussion on the routing policies in ISP
networks. Policies are usually translated to a set
of rules in a particular order of precedence, and
are used to determine the preference of one route
over the other. Such policies can be incorporated 
in defining the weights of the edges of $\mathcal{R}_x$,
and/or in the process of computing the paths in 
$\mathcal{R}_x$. In the extreme case (when an AS
does not wish to share its policy-based route selection
rules with its neighbors), information about the 
graph $\mathcal{R}_x$ can be retrieved by each
node $x_i$ from $x$, in order to construct $\mathcal{R}_x$
locally, in order to compute its own alternate
path to $s$. Note that since the average degree of
a node is usually small (within 20-40), the size
of $\mathcal{R}_x$ would typically be reasonably small.

\section{Concluding Remarks}
\label{concl}

In this paper we have presented an efficient distributed algorithm for
the computing alternate paths that avoid a failed node. To the best
of our knowledge, this is the first completely decentralized algorithm
that computes such alternate paths. All previous algorithms, 
including those presented in \cite{bg04jgaa,bg08snfr,lynzc,sl,wg08,znylwc} 
are centralized algorithms that work using the information of the 
entire network graph as input to the algorithms.

The paths computed by our algorithm are required by the single node
failure recovery protocol of \cite{bg08snfr}. They also qualify as the
{\em reverse paths} required by the BRAP protocol of \cite{wg08}, 
which deals with single link failure recovery. Our distributed
algorithm computes the exact same paths as those generated by
the centralized algorithm of \cite{bg08snfr}, and even though
not optimal alternate paths, they are usually good - within
$15\%$ of the optimal for randomly generated graphs with 
$100$ to $1000$ nodes, and with an average node degree of 
upto $35$. The reader is referred to \cite{bg08snfr} for
further details about the simulations.

Our algorithm can be generalized to solve 
other similar problems. In particular, we can derive 
distributed algorithms for the single link failure
recovery problem \cite{bg04jgaa,sl}, the minimum spanning
tree sensitivity problem \cite{dixon}, and the detour-critical
edge problem \cite{npw-detour}. The cited papers 
present centralized algorithms for the problems studied. 
All these are link failure recovery problems that
deal with the failure of one link at a time.
In these problems, for each tree edge (minimum spanning
tree, or shortest paths tree, depending on the problem), 
one needs to find an edge across the cut induced by the deletion 
of the edge. We essentially need to find edges similar to 
the green edges for the SNFR problem, except for one minor
change: these green edges have one end point in the node's subtree,
and the other outside its subtree (for the SNFR problem, the other
end point needs to be outside the subtree of the node's parent).
Our DFS labeling scheme can be used for determining whether
an edge is green or not according to this definition. Using 
the DFS label computation algorithms of \cite{fppp}, and our 
protocols for edge propagation ({\tt RecordNonTreeEdge}), 
we can find the required alternate paths that avoid a 
failed edge. 

We believe that our techniques can be
generalized to solve some other problems as well.

In their recent work, Kvalbein, et. al. \cite{kcg07} address 
the issue of load balancing when a proactive recovery scheme
is used. While some previous papers have
also investigated the issue, as mentioned in \cite{kcg07}, they
usually had to compromise on the performance in the failure-free
case. To a somewhat limited extent, our algorithm can be modified
to take this aspect into consideration. For instance, instead of
computing the shortest paths tree $\mathcal{T}_{s_x}$ 
in $\mathcal{R}_x$, one is free to compute 
other types of paths from each node $y_i$ to $s_x$ in order 
to ensure that the same set of edges don't get used in 
many recovery paths.

\bibliography{dsnfr}
\bibliographystyle{plain}

\end{document}